\documentclass[aip,jap,amsmath,amssymb,reprint,superscriptaddress,floatfix]{revtex4-1}
\usepackage{graphicx}%
\usepackage{dcolumn}%
\usepackage{bm}%
\usepackage{epstopdf}
\usepackage[hyperindex,breaklinks]{hyperref}
\usepackage[caption=false]{subfig}

\newcommand{\zr}{\mbox{Zr}}
\newcommand{\zc}{\mbox{Cu}_{50}\mbox{Zr}_{50}}
\newcommand{\zca}{\mbox{Cu}_{50}\mbox{Zr}_{45}\mbox{Al}_{5}}

\begin{document}

\title{Assessing the Reliability of Minimally Constrained Reverse Monte Carlo Simulations for Model Metallic Liquids}
\date{\today}
\author{R. Ashcraft}
\affiliation{Department of Physics, Washington University in St. Louis, St. Louis, Missouri, 63130, USA}
\author{K. F. Kelton}
\email[Author to whom correspondence should be addressed.]{kfk@wustl.edu}
\affiliation{Department of Physics, Washington University in St. Louis, St. Louis, Missouri, 63130, USA}
\affiliation{Institute of Materials Science and Engineering, Washington University in St. Louis, St. Louis, Missouri, 63130, USA}
\begin{abstract}
Molecular dynamics simulations using semi-empirical potentials are examined for three liquids to check the reliability of reverse Monte Carlo (RMC) simulations to reproduce atomic configurations when only total pair correlation functions (TPCF) are used as constraints.  The local structures are determined from a Voronoi tessellation of the ensemble and compared with the structures obtained by RMC in terms of asphericity, volume, coordination number, Voronoi index, and nearest-neighbor distance.  It is found that in general the distributions generated from RMC do not match the MD configurations, using the $L^1$ (taxicab) distance as a metric, although in some cases a measure of central tendency for the distribution did match. Since only TPCFs are typically used to constrain the RMC simulations of experimental data, this study establishes the limits on what can be learned by this analysis. It indicates that caution should be used when interpreting RMC-generated structures using few constraints since many structural quantities are not reproduced well.
\end{abstract}

\maketitle

\section{Introduction}\label{intro}
The Reverse Monte Carlo (RMC) method~\cite{McGreevy1988,McGreevy2001} is a common technique used to obtain 3D atomic structures for liquids and glasses using data obtained rom X-ray diffraction or neutron scattering measurements. In this method, atom positions in an atomic ensemble are adjusted using a Monte Carlo algorithm to give the best match to the total structure factor (TSF) or total pair correlation function (TPCF) derived from the scattering data. For an alloy containing $n$ elements, the typical number of structural constraints necessary is $n(n+1)/2$. In practice, however, it is generally difficult to experimentally measure these, particularly for samples containing more than two elements. Instead the information is frequently obtained from \emph{ab. initio}~\cite{Car1985} or molecular dynamics (MD) simulations and is typically in the form of partial pair correlation functions (PPCFs)~\cite{Wang2008a}. Attempts have also been made for a combined procdure, simulating the system using MD then performing RMC with experimental data, or incroporating the interatomic potential functions in the RMC code~\cite{Gereben2015}. Often, however, RMC fits are made to only the TSF or TPCF. These fits will give the most random structure consistent with the experimental scattering data and will lack elementally resolved information. The validity of this approach is therefore questionable.

	Previous studies have examined the reliability of the RMC method for MD-generated structures for elemental~\cite{Smolin2002,Gelchinski1999} and binary~\cite{Mendelev2010,Fang2014,Almyras2011} systems, typically constraining the fit with all the measured PPCFs. The results of these studies typically suggest that the RMC method can accurately recreate the atomic structure of the the system. This is based on an examination of the RMC structure using metrics including Voronoi tessellation~\cite{Voronoi1908a,Finney1970}(the most common), bond angle distribution~\cite{Cheng2011}, and Honeycutt-Anderson analysis\cite{Honeycutt1987}. However, in at least one study it is found that the RMC generated structure can be considerably more disordered than the MD-generated one.
	
	Here, the case most often used for the analysis of experimental data will be considered, i.e. where the RMC is constrained with only the TPCF, termed hear as a minimally constrained reverse Monte Carlo (mcRMC) simulation. The reliability of the atomic structures generated by these mcRMC simulations and how that reliability depends on the temperature and the number of elements are examined. To examine the latter, three systems ($\zr$, $\zc$, and $\zca$) are simulated at several temperatures using classical MD. The TPCF is calculated from each simulation and used as input to mcRMC simulations. The atomic configurations from both the mcRMC and MD simulations are then compared using Voronoi tessellation.
	
\section{Simulations and Analysis Methods}\label{methods}
\subsection{Molecular Dynamics Simulations}
The TPCFs were obtained from MD simulations using the LAMMPS~\cite{Plimpton1995a} software employing embedded atom method~\cite{Cheng2011} (EAM) potentials for $\zr$ ~\cite{Mendelev2007}, $\zc$ ~\cite{Mendelev2009}, and $\zca$ ~\cite{Cheng2009}. All compositions were simulated with 15,000 atoms under the NPT ($P=0$) ensemble with periodic boundary conditions. The Nos\'e -Hoover thermostat~\cite{Nose1984a,Hoover1985} was used to equilibrate each system at each target temperature before data collection. The atomic configuration for each system was randomly initialized and evolved in the high temperature liquid before cooling ($3-8 \times 10^{11}$K/s) to each target temperature. To reach equilibrium the system was evolved for $5-15$ns.($3 \times 10^6$ MD time steps), depending on the values of the temperature and composition. Each PPCF, $g_{\alpha \beta}(r)$,  was then calculated by averaging over 15,000 snapshots of the system using (see~\cite{Cheng2011})
\begin{equation}
	\label{eq:ppcf}
	g_{\alpha \beta}(r) = \frac{N}{4 \pi r^2 \rho N_{\alpha} N_{\beta}} \sum_{i,j=1}^{N_{\alpha},N_{\beta}} \delta (r-|\mathbf{r}_{ij}|)
\end{equation}
where $N$ is the number of atoms, $\rho$ is the number density, $|\mathbf{r}_{ij}|$ is the distance from atom $i$ to atom $j$, and $N_{\alpha}$ and $N_{\beta}$ are the number of $\alpha$ and $\beta$ atoms, respectively. The TPCF was calculated within the Faber-Ziman~\cite{Waseda1980} formalism 
\begin{equation}
	\label{eq:tpcf}
	g(r) = \sum_{\alpha}\sum_{\beta} \frac{c_{\alpha} c_{\beta} b_{\alpha} b_{\beta}}{\left\langle b \right\rangle^2} g_{{\alpha}{\beta}}(r)
\end{equation}
where $c_{\alpha}$ is the atomic concentration and $b_{\alpha}$ is the neutron scattering length for element $\alpha$ and $g_{\alpha\beta}$ is the PPCF between elements $\alpha$ and $\beta$. While the case of neutron scattering is assumed for the analysis presented here, the approach could be directly extended to X-ray scattering if $q$-dependent atomic scattering factors were used

The viscosity of each liquid was calculated using the Green-Kubo formula~\cite{Hansen2013}. The viscosity exhibits a crossover from near-Arrhenius to super-Arrhenius behavior at the temperature $T_A$. Since it is difficult to calculate the melting temperature from MD simulations, $T_A$ was used as the scaling temperature, because it is readily computed.

\subsection{Reverse Monte Carlo Simulations}
	As discussed, minimally constrained reverse Monte Carlo simulations are carried out by taking an input configuration of atoms and input TPCF, changing the input configuration by randomly moving an atom in a random direction and then computing the mcRMC values of the TPCF for this new configuration. This is then compared with the MD-generated TPCF using the $\chi^2$ as a measure of the goodness of fit,
\begin{equation}
\chi^2 = \sum_i \frac{\left[ g^{\mathrm{RMC}}(r_i)-g(r_i) \right]^2}{\sigma^2}
\end{equation}
$\sigma$ is the reliability of the data set. The random move is accepted if the $\chi^2$ is reduced and is accepted with a Boltzmann probability if the $\chi^2$ is increased. This procedure is repeated until the $\chi^2$ is minimized. 

	The mcRMC simulations were run using the RMC++~\cite{Gereben2007a} software. Each mcRMC simulation started with a random configuration of 10,000 atoms confined to a cubic box with periodic boundary conditions, with the box dimensions consistent with the number density predicted from the MD results.  The hard-sphere cutoff distances were determined from the value of $r$ where each respective PPCF trended to 0 on the low-$r$ side of the main peak. Ten separate simulations were run at each temperature to generate more reliable distributions from the Voronoi tessellation. Each simulation was run in parallel for 30-45 computational hours depending on the resulting $\chi^2$. Convergence of the mcRMC was assumed when both the value and the change with time of the $\chi^2$ was suitably small ($\chi^2 < 10$).

\subsection{Voronoi Tessellation}\label{ss:vt}
The Voronoi tessellation divides the atomic configuration into Voronoi polyhderal (VP) cells, each consisting of a central atom and the space closer to this atom than any other. The VP is constructed as the collection of perpendicular bisecting planes between the central atom and all neighboring atoms; the planes form the faces of the VP. It has been shown~\cite{Park2012} that for atoms of different radii the standard Voronoi tessellation technique can lead to significant errors. The radical Voronoi tessellation~\cite{Gellatly1982,Gerstein1995} was therefore used for the $\zc$ and $\zca$ simulations. This technique weights the placement of the bisecting planes by the radii of the central and neighbor atoms. Another potential error that is common in Voronoi tessellation is the occurrence of exceptionally small faces and edges, which arise from more distant atoms~\cite{Brostow1998}. The effect of removing these faces and edges is not currently investigated but may lead to broader distributions for nearest-neighbor distance and coordination number in both the MD and mcRMC.

The Voronoi tessellation was carried out using a Python extension of the VORO++\cite{Rycroft2006,Rycroft2009} code. The Goldschmidt radii~\cite{Gale2004} were used for the radical Voronoi tessellation. The VP can be described by the Voronoi index (VI), $\langle n_3,n_4,n_5, \ldots \rangle$, where $n_i$ is the number of $i$-edged faces. Both the geometric coordination number (CN), number of faces, and the nearest-neighbor distance (NND), distance between atoms that share faces, were calculated. Finally, the volume and surface area of each VP were used to calculate the asphericity parameter, $\alpha = S^3/(36 \pi V^2)$, which gives a measure of how similar the VP is to a sphere ($\alpha=1$) or a given regular polyhedron ($\alpha = 1.32503$ for a regular dodecahedron).  

\subsection{$L^{1}$ Histogram Distance}
The $L^{1}$, or Taxicab/Manhattan, distance was used to compare the similarity of the distributions obtained from the Voronoi tessellation. Other metrics such as the Bhattacharyya\cite{Bhattacharyya1943} or Hellinger~\cite{Hellinger1909} distance give similar but systematically larger results. The $L^{1}$ distance is given by
\begin{equation}
L^{1}(\textbf{X},\textbf{Y}) = \frac{1}{2}\sum_{i}^{n} \left| x_i-y_i \right|
\label{eq:L1}
\end{equation}
where $\textbf{X}$ and $\textbf{Y}$ are probability vectors (i.e. $\sum x_i = 1$) and the factor of $1/2$ is included for normalization. By definition it is easy to see $L^{1}=0$ for identical distributions and $L^{1}=1$ for distributions that have no overlap. This metric, therefore, gives a unique measure on the reliability of RMC to reproduce distributions from the MD atomic structure.

\section{Results and Discussions}\label{results}

\begin{figure}
	\includegraphics[width=\linewidth]{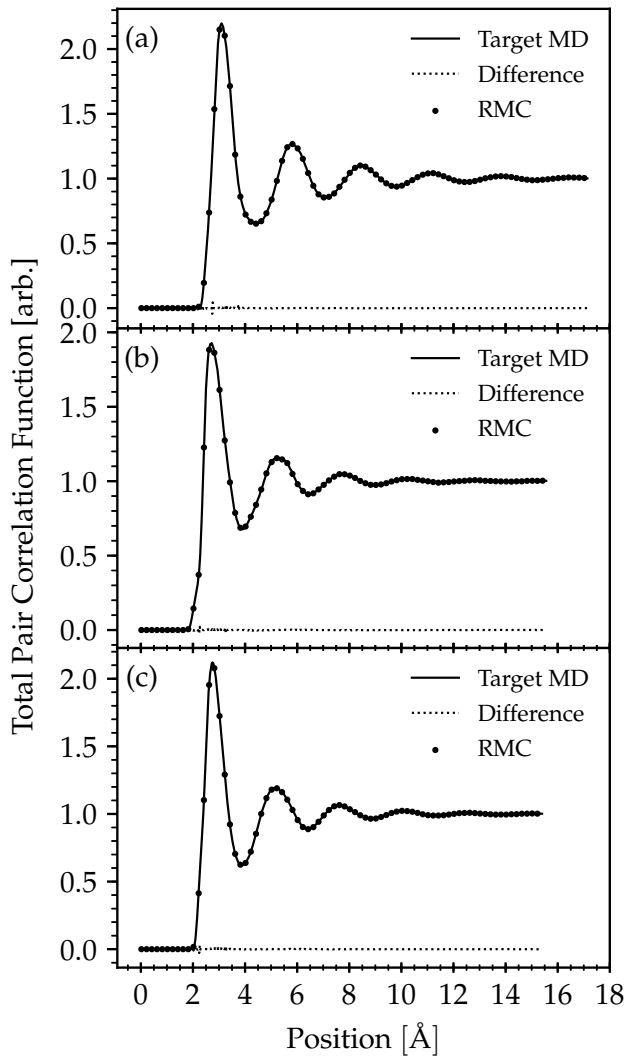}
	\caption{\label{fig:RMCfit} Representative plots of the RMC fit (circles) to MD data (line) for (a) $\zr$ , (b) $\zc$ , and (c) $\zca$  at 2500, 2000 and 1500K, respectively. The difference curve (dashed line) for each data set is also included to emphasize the level of agreement.}
\end{figure}

	The mcRMC fits (averaged over ten simulations) are compared in Fig~\ref{fig:RMCfit} with the MD TPCFs for representative temperatures of each composition. The fits are extremely good, indicating that the generated structures should be a good approximations to the atomic configurations in the MD data. However, as mentioned, chemical ordering was not reproduced well for both $\zc$ and $\zca$ because only a single constraint was used. Only examples at intermediate temperatures are shown since the $\chi^2$ for each composition appears to be temperature dependent. This is likely due to the random nature of the RMC algorithm; the RMC algorithm gives the most disordered configuration consistent with the input constraints. Since the high temperature data are inherently more disordered they are easier to fit than the lower temperature data resulting in a lower $\chi^2$, although the variation in the $\chi^2$ with temperature is relatively small.
	
\begin{figure}
	\includegraphics[width=\linewidth]{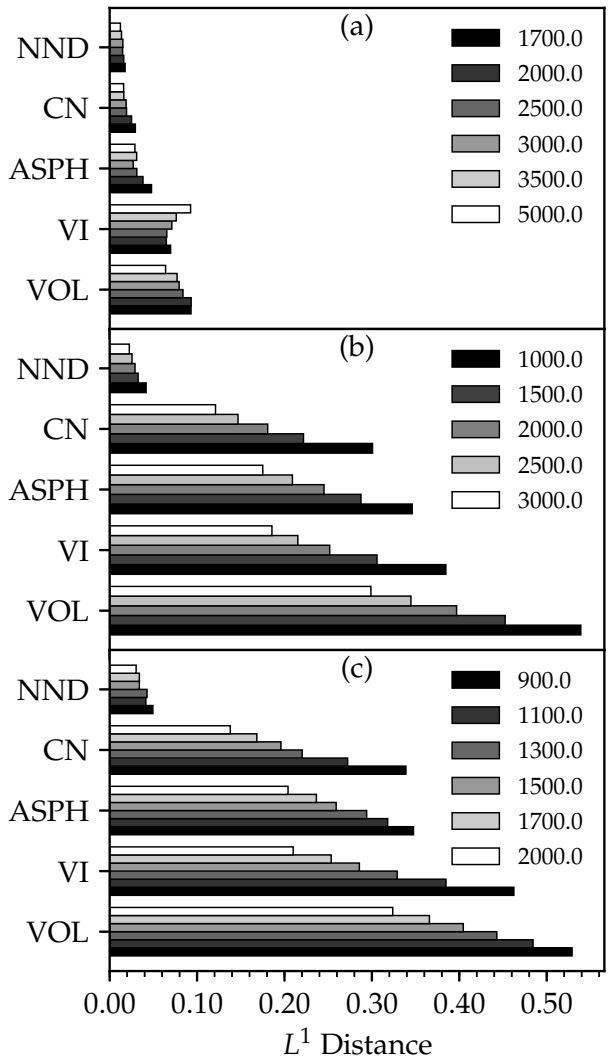}
	\caption{\label{fig:L1Dist} $L^1$ (Manhattan) distance between mcRMC and MD distributions of nearest neighbor distance (NND), coordination number(CN), asphericity parameter (ASPH), Voronoi index (VI), and volume (VOL) for (a) $\zr$,(b) $\zc$, and (c) $\zca$. The darkness of the shading indicates the temperature $[K]$ where black is low and white is high. The $L^1$ distance is normalized between 0 and 1 where 0 indicates two identical distributions and 1 indicates no overlap in the distributions.}
\end{figure}	
	
\begin{figure*}
	\includegraphics[width=\linewidth]{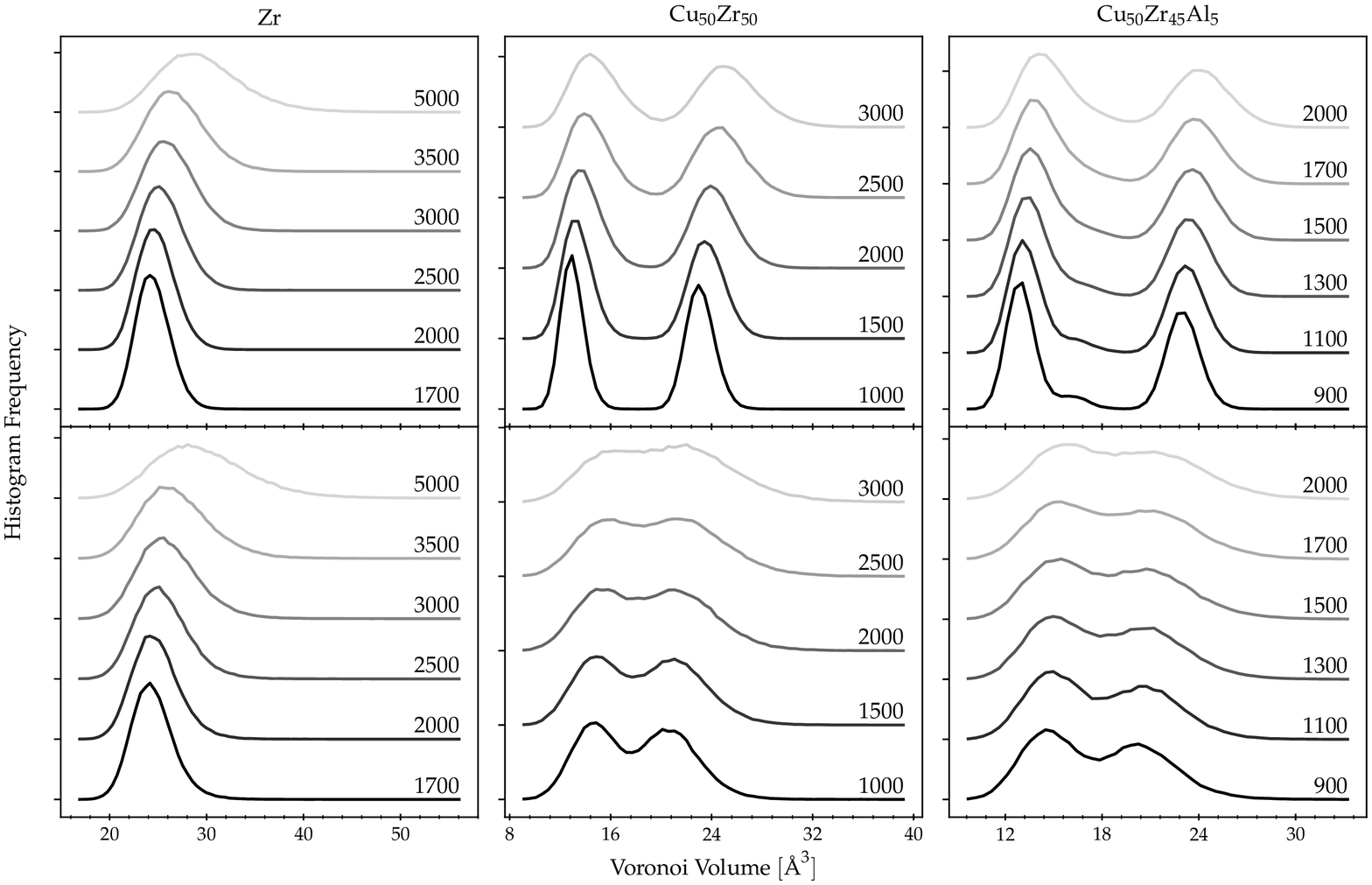}
	\caption{\label{fig:Vol} Frequency of Voronoi volume from MD (top row) and mcRMC (bottom row) for each composition. The temperatures [K] are indicated by shading (light is high and dark is low) and indicated on the right side of each curve. Each distribution is offset for clarity. Note that the distributions for mcRMC and MD match well for $\zr$ but not for $\zc$ and $\zca$. Higher temperature distributions are broader for both mcRMC and MD.}
\end{figure*}	
	
	The $L^1$ distance for each of the previously discussed parameters obtained from the Voronoi tessellation procedure are shown in Fig.~\ref{fig:L1Dist}. These each give in a general sense a property of the VP associated with the tessellation. By comparing their values between RMC and MD, they can give an indication of reliability. The Voronoi volume gives the general size of the space that is allocated to each atom. The asphericity gives information about the general distribution of atoms about the central atom by considering its size. Both the coordination number and Voronoi index give similar information about the distribution, but irrespective of geometric size and with varying degrees of sensitivity. Lastly, the nearest-neighbor distance gives information on the distance between the polyhedra.
	
\subsection{Volume}
	The Voronoi volume is arguably one of the most easily reproducible properties considered here since it is determined by the input number density, simulation volume, and hard sphere radii. However, because the mcRMC simulation is constrained solely with the TPCF it is unable to accurately allocate the proper volume to each element. This is shown in Fig~\ref{fig:Vol}, where the Voronoi volume distribution is shown for each composition and temperature. While the $\zr$ liquid is reproduced well the alloy liquids are not. The $\zc$ and $\zca$ liquids have two and three peaks, respectively, one for each element. These peaks are distinct in the structure obtained from the MD but are broadened and overlapping for the mcRMC structure. This difference, which is directly related to the use of only the TPCF constraint, is the cause of the large $L^1$ distance observed in Fig.~\ref{fig:L1Dist} and the unreliability of the Voronoi volume in minimally constrained systems. However, if the Voronoi tessellation fills space the average Voronoi volume should be the reciprocal of the number density indicating that the average volume should still be a reliable parameter, as shown in Fig.~\ref{fig:Vol}. While the details of the atomic distribution might be unreliable the average properties of the distribution could still be useful.

\begin{figure}
	\includegraphics[width=\linewidth]{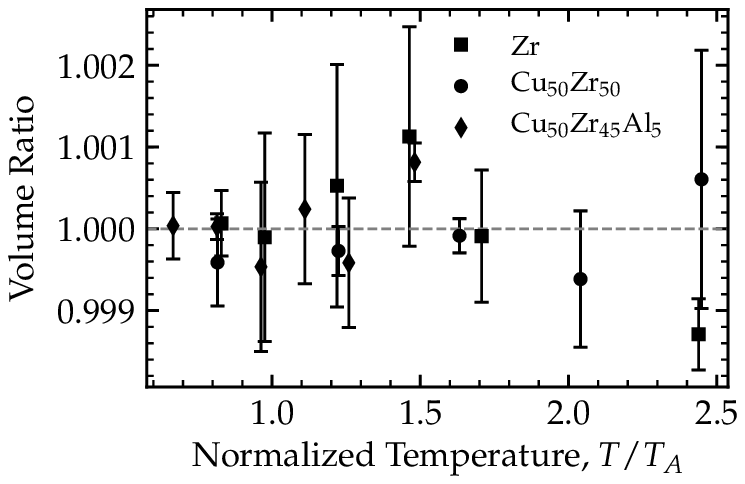}
	\caption{\label{fig:VolRatio} Ratio (MD/mcRMC) of average Voronoi polyhedra volume for $\zr$ (square), $\zc$ (circle), and $\zca$ (diamond) versus temperature, normalized to the Arrhenius crossover temperature $T_A$. This ratio should be equal to one (dashed gray line).}
\end{figure}

\subsection{Asphericity}
\begin{figure}
	\includegraphics[width=\linewidth]{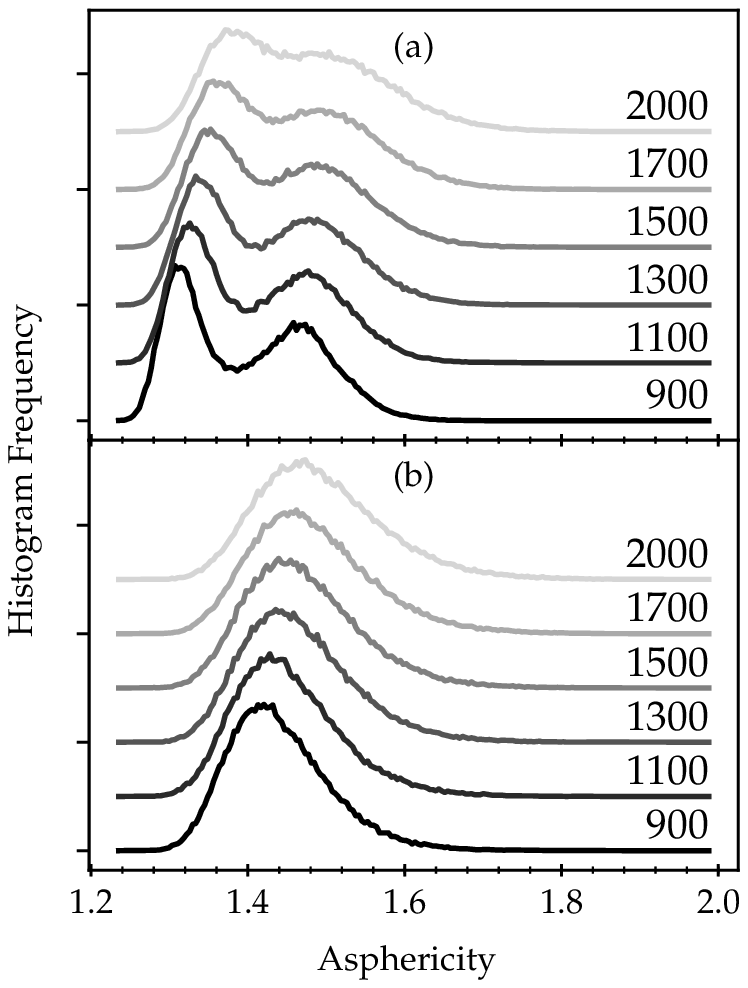}
	\caption{\label{fig:Asph} Histogram frequency for asphericity for $\zc$ from (a) MD and (b) mcRMC. Each line indicates a different temperature; the curves are offset for clarity.}
\end{figure}
	Asphericity has been used to examine the shape of the VP in liquid water, where it was noted that the volume is not correlated with $\alpha$ and that it approaches the value of ice upon cooling the liquid~\cite{Sampoli1991}. In a similar fashion asphericity is used here to see how the shape of the VP change with cooling and how well the mcRMC recreates the shape of the VPs. As shown in Fig.~\ref{fig:Asph} for $\zc$ there are two distinct peaks in the MD (a) compared to the mcRMC (b). This is a consequence of failing to resolve the elemental differences, since the two elements have significantly different distributions, and results in the relatively large $L^1$ distance (Fig.~\ref{fig:L1Dist}). Only the data for $\zc$ are shown here but similar trends are found in the other liquids. As observed in Fig.~\ref{fig:Asph}, the majority of the distribution is in the region $1.25<\alpha<1.75$ for both the MD and mcRMC data. This corresponds to shapes that are close to that of a regular dodecahedron ($\alpha=1.325$) but the regular octahedron ($\alpha=1.654$) is also present. While the mcRMC fails to recreate the elementally resolved properties, worse than for the volume distribution, it does give the correct bounds for the asphericity. 
	
\begin{figure}
	\includegraphics[width=\linewidth]{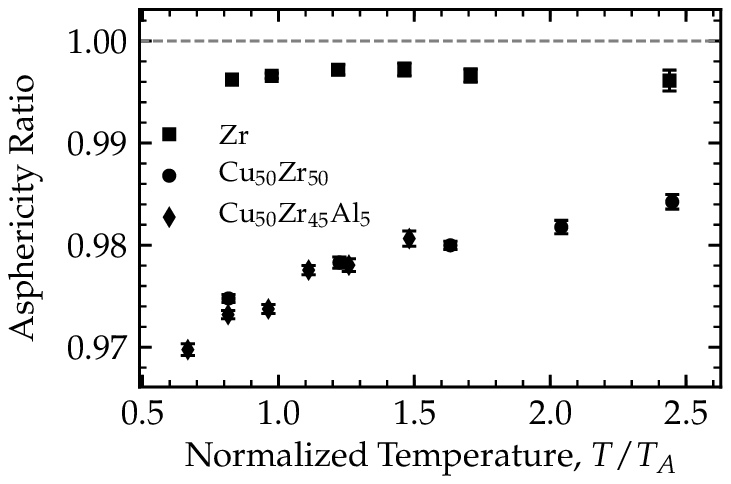}
	\caption{\label{fig:AsphRatio} Ratio (MD/mcRMC) of the asphericity parameter for $\zr$ (square), $\zc$ (circle), and $\zca$ (diamond) versus temperature normalized to the Arrhenius crossover temperature $T_A$. If the mcRMC analysis recreated the MD configurations perfectly the ratio should be one (dashed line).}
\end{figure}

	Upon cooling the average asphericity decreases toward the value of the regular dodecahedron (and other similar polyhedra). This is the case for all liquids studied because the magnitudes of their asphericities are similar. The ratio of the average MD to average mcRMC values (Fig.\ref{fig:AsphRatio}) shows that the mcRMC fails to recreate the average properties of the VP shape distribution, both in magnitude and temperature dependence, except for $\zr$ where only the magnitude is marginally different. This result for $\zr$ is likely due to the increased disorder from the mcRMC simulation.  

\subsection{Voronoi Index and Coordination Number}
\begin{figure}
	\includegraphics[width=\linewidth]{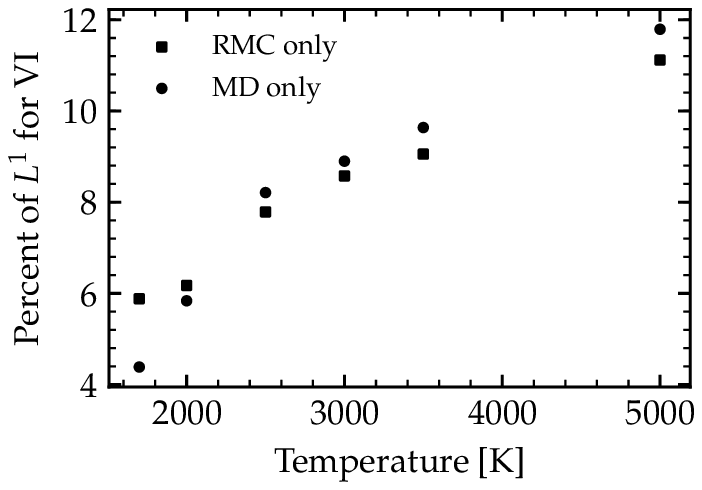}
	\caption{\label{fig:L1VI} The percent of the Voronoi index $L^1$ distance for $\zr$ as a function of temperature that is due to polyhedra only in RMC (square) and only in MD (circle). }
\end{figure}

	The VI is perhaps the most commonly used parameter to describe the local environment from a Voronoi tessellation. It is also the least general of the parameters discussed here, since it describes average properties of groups of VPs rather than system-wide average properties. Ash shown earlier in Fig~\ref{fig:L1Dist}, the VI has a relatively large $L^1$ distance for all compositions. The $\zr$ VI $L^1$ distance is, however, surprising due to the anomalous temperature dependence. It is also interesting to note that even for $\zr$ the relatively large $L^1$ distance of the VI indicates that the population of VIs do not properly reflect that in the MD data, even with a fully constrained system. One contributing factor to this anomalous increase is the increase of VPs that are only present in either the mcRMC or the MD results, which in general increases with temperature. The total percentage attributable to these factors at high temperatures can be as large as $\%20$ (Fig.~\ref{fig:L1VI}). However, for $\zr$ the temperature dependence of these factors is not enough to change the temperature dependence, which points to another factor. Whether this is a true effect or is an artifact is still under investigation. 
	
\begin{figure}
	\includegraphics[width=\linewidth]{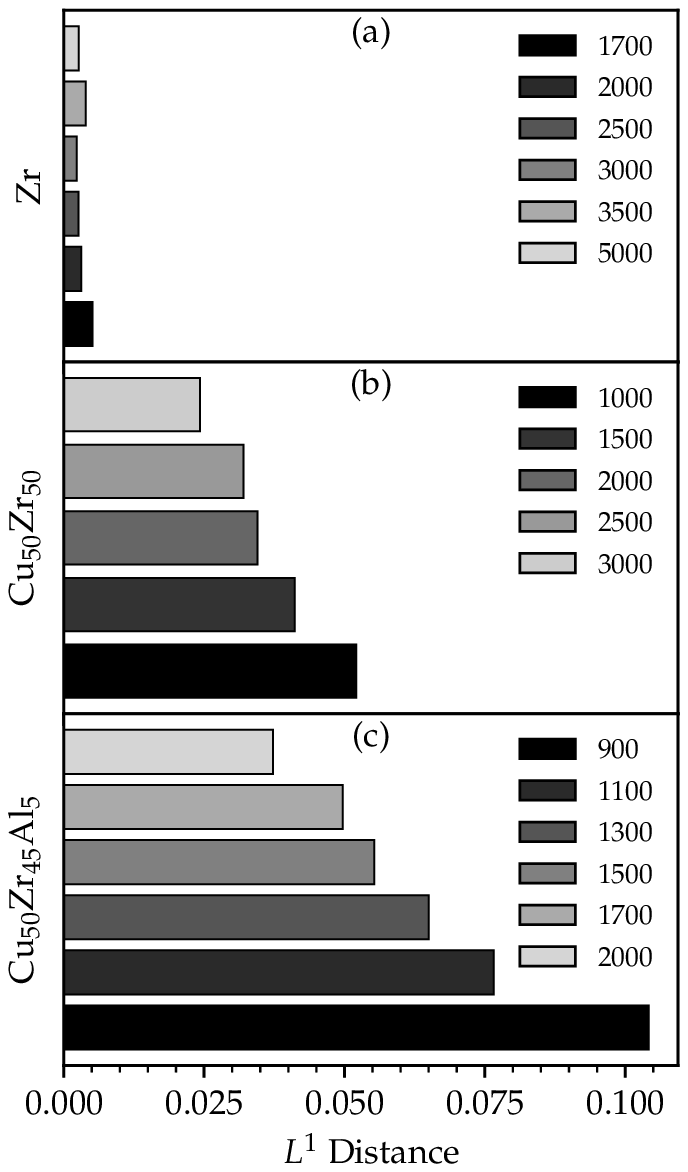}
	\caption{\label{fig:VIAve} $L^1$ distance for the average VP of (a) $\zr$, (b) $\zc$, and (c) $\zca$. The average VP is computed by extending the typical VI definition to allow for fractional faces.}
\end{figure}

	The use of any single VI to determine agreement between the methods is difficult since different systems will in general prefer different structures. If the VI definition is extended to allow for fractional numbers of faces an "average" VP can be constructed for each simulation. No physical meaning is attached to the fractional numbers of faces; rather it provides another measure of mcRMC reproducibility of the atomic environment. Performing this analysis on the three compositions studied here and then comparing the average VPs from mcRMC and MD gives extraordinarily low $L^1$ distances (Fig.\ref{fig:VIAve}) that are in-line with the values for the NND (Fig.~\ref{fig:L1Dist}). In this sense the average atomic environment is reproduced quite well from mcRMC. However, a distinct dependence on the number of elements is observed, indicating that while the average atomic environment is reproduced well the addition of more types of atoms, even in small amounts, decreases the reliability of the mcRMC. This is not surprising since the amount of missing information for the mcRMC increases with the addition of more elements. To our knowledge, however, this analysis is the only one that shows such clear evidence of this effect.
	
	The CN is included because of its strong relation to the VI. However, it suffers from the same issues as the volume and many other parameters. The mcRMC is unable to allocate the proper space for different types of atoms, causing the distribution of CNs to lie somewhere between the elemental distributions. This is also the cause of the relatively large $L^1$ distance as shown in Fig.~\ref{fig:L1Dist}. 
	
\subsection{Nearest-neighbor Distance}
\begin{figure}
	\includegraphics[width=\linewidth]{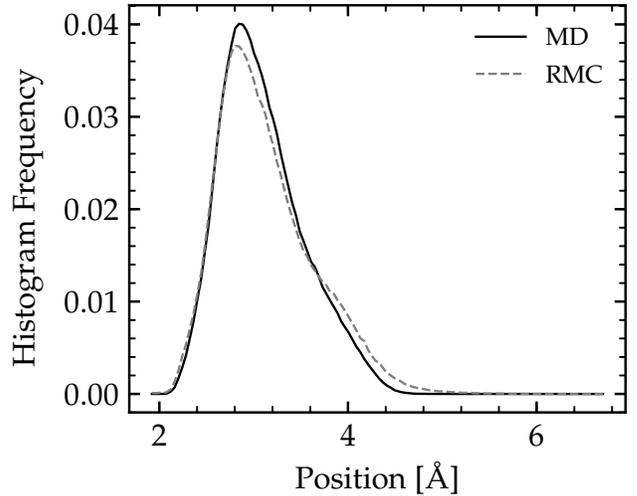}
	\caption{\label{fig:NNDist} Nearest-neighbor distance histograms for both the  MD (solid line) and mcRMC (dashed line) simulations of $\zca$ at $T=900$K. The histogram frequency is normalized so that it sums to one. Only the lowest temperature distribution for $\zca$ is shown because it has the largest differences for all of the compositions and temperatures examined.}
\end{figure}

In addition to the volume the NND is the other most easily reproducible parameter from RMC, since the input TPCF data is inherently related to this distance through the radial distribution function. Even when using the TPCF as a single constraint the RMC simulation should give a reliable distribution for the NND on average, although chemical effects and elementally resolved distances will still be absent. This reliability is reflected in the low $L^1$ distance shown in Fig.~\ref{fig:L1Dist}. The $L^1$ distance is the smallest of any parameter considered regardless of the number of elements or temperature. As shown in Fig.~\ref{fig:NNDist} good agreement is found between MD and mcRMC, even for the worst case examined. Reverse Monte Carlo simulations tend to place atoms at marginally larger distances compared to the MD-generated structures, which is a consequence of the lack of input elemental and chemical information. However, the increased distance is asymmetric, since the low distance side is constrained by the hard-sphere cutoffs. 

\begin{figure}
	\includegraphics[width=\linewidth]{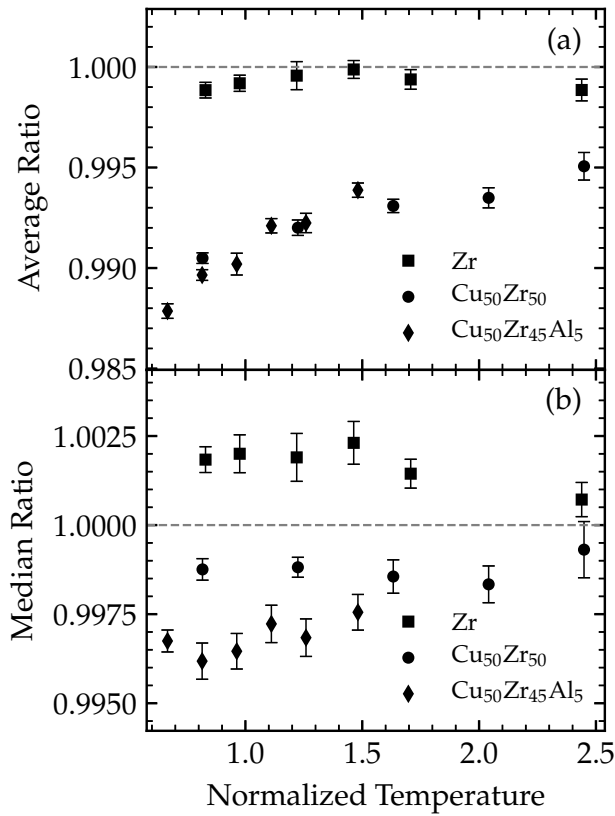}
	\caption{\label{fig:BL} Plot of the MD to mcRMC ratio of average (a) and median (b) nearest-neighbor distances for $\zr$ (square), $\zc$ (circle), and $\zca$ (diamond) versus temperature normalized to $T_A$, the Arrhenius crossover temperature. The ideal value of one is marked by a dashed gray line.}
\end{figure}

	Since the $L^1$ distance is small the average atomic position (bond length) should be well characterized by the RMC simulation. However, due to the asymmetric change in distance the bond length will have a modified temperature dependence compared to that from the MD simulation. A possible solution to this is to instead consider the median bond length as a measure of central tendency, which is less affected by this increased asymmetry. The ratios of MD to mcRMC ratio for both the median and mean NND are shown in Fig.~\ref{fig:BL} for all compositions. In Fig.~\ref{fig:BL}a the average NND for $\zc$ and $\zca$ shows a different temperature dependence between the mcRMC and MD results. The effect of the RMC placing atoms at larger distances at lower temperatures is to depress the actual change which can be seen in the broadening of the first peak in TPCFs from scattering studies~\cite{Ding2014}. Furthermore, the median NND (Fig.~\ref{fig:BL}b) shows relatively no temperature dependence for all compositions and a slightly reduced amplitude in comparison to the average value. This would indicate that even with no other constraints the NND distribution and even more so its median are reliable.

\section{Summary}\label{conclusion}
	In the present study atomic structures created from molecular dynamics (MD) simulations for three different liquids were used to explore how well reverse Monte Carlo (RMC) fits to the MD-generated pair correlation functions reproduce the atomic structures. The worst case (also generally the one used to fit experimental data) was examined, using only the measured total pair correlation function to constrain the RMC fits (termed here an mcRMC analysis). An in-depth analysis of the Voronoi tessellation for the structures obtained from the MD and mcRMC was made by examining the distributions and measures of central tendency of the Voronoi volume, asphericity, Voronoi index, coordination number, and nearest-neighbor distance using the $L^1$ distance as a metric of similarity. While the mcRMC is able to reproduce some properties of each distribution, the structures were generally not well reproduced. That the fit properties were questionable raises doubts about the use of RMC for systems that are not fully constrained and the validity of more demanding properties of the configuration (i.e. network analysis). However, the predicted mean volume and median nearest neighbor distances were better predicted, indicating that mcRMC can be reliably used to analyze experimental data to obtain these quantities. The temperature dependence of the distribution similarities was also examined. In general even for the fully constrained $\zr$ liquid, the $L^1$ distance increases as the temperature decreases, indicating that the mcRMC results become less reliable. Extending these results to multicomponent systems such as metallic glasses, which are more likely to be constraint deficient, calls into question the reliability of mcRMC results. Further the glasses are generally more ordered than the liquids. Since the structure obtained from RMC simulations are the most disordered ones consistent with the scattering data, it is not possible to obtain a clear picture of the order except for in an averaged sense.

\section*{Acknowledgments}
The work at Washington University in St. Louis was partially supported by the National Science Foundation under Grant DMR-12-06707.

\bibliographystyle{apsrev4-1}
\bibliography{library}

\end{document}